\documentclass[pra,twocolumn,preprintnumbers,amsmath,amssymb,nofootinbib,showpacs]{revtex4-1}
\usepackage{graphicx}
\usepackage{amssymb}
\usepackage{mathrsfs}
\usepackage{color}

\newcommand{\beq}{\begin{equation}}
\newcommand{\eeq}{\end{equation}}

\begin{document}

\title{Reply to ``Comment on ``Galilean invariance at quantum Hall edge''"}

\author{Sergej Moroz$^{1,2,3}$, Carlos Hoyos$^4$ and Leo Radzihovsky$^{2,3}$}
\affiliation{$^1$Physik-Department, Technische Universit\"at M\"unchen, D-85748 Garching, Germany \\
                $^2$Department of Physics, University of Colorado, Boulder, Colorado 80309, USA \\
                $^3$Center for Theory of Quantum Matter, University of Colorado, Boulder, Colorado 80309, USA \\
                $^4$Department of Physics, Universidad de Oviedo, Avda. Calvo Sotelo 18, 33007, Oviedo, Spain}

\begin{abstract}
Motivated by a recent Comment by J. H\"oller and N. Read [Phys. Rev. B 93, 197401 (2016)], we revisit the problem of a chiral Luttinger liquid on a boundary of a Galilean-invariant quantum Hall fluid. After correcting the linear response calculation, the real part of the longitudinal conductivity derived in the model constructed in our paper [Phys. Rev. B 91, 195409 (2015)] agrees with the result found in the Comment for non-interacting fermions confined by a linear potential. We also withdraw our previous conjecture that the longitudinal conductivity contains a universal contribution determined by the  ``shift'' and provide arguments demonstrating its non-universal nature. 
\end{abstract}


\maketitle

In a recent paper \cite{Moroz2015} we constructed a theory of a chiral Luttinger liquid on a boundary of a Galilean invariant quantum Hall fluid. Using this theory we considered electromagnetic response at the edge and computed the longitudinal electric conductivity in the low frequency and small wave vector regime
\beq \label{ourcond}
\sigma(\omega, p_x) = \frac{\nu}{2 \pi} \Big( 1 +\frac{\mathcal{S}}{4} \frac{p^2_x}{B} \Big) \frac{i c }{\omega- c p_x+i0^+} +im \epsilon''(B)\frac{p_x}{B},
\eeq
where we introduced the filling fraction $\nu$, the magnetic field  $B>0$,  the velocity $c$ of the chiral edge mode and the particle mass $m$. Note that the definition of the conductivity in this Reply differs from the (unconventional) one used in our original paper \cite{Moroz2015} by an overall sign. This choice ensures that the real part of the conductivity is positive. The conductivity \eqref{ourcond} differs from the conductivity extracted from the well-known edge theory of Wen \cite{Wen1991a, Wen1992a} and Stone \cite{Stone1991} by higher derivative corrections that depend on the ``shift'' $\mathcal{S}$ introduced in \cite{Wen1992}  and the second derivative of the energy density $\epsilon(B)$ of the quantum Hall fluid. In our work we proposed that the first term in Eq. \eqref{ourcond} is universal and suggested that a spectroscopic measurement of the longitudinal conductivity might be a new way to measure the ``shift'' $\mathcal{S}$ and the closely related Hall viscosity.

In a recent Comment \cite{Holler2016} on our paper, H\"oller and Read computed the spectral weight (the real part of the longitudinal conductivity integrated over the frequency $\omega$) for the integer quantum Hall system of noninteracting fermions that occupy any number of the lowest Landau levels in the presence of an edge.  For a linear confining potential, they found a result consistent with our prediction \eqref{ourcond} except for a distinct numerical coefficient in front of the  ``shift'' $\mathcal{S}$. They also demonstrated that the spectral weight depends on the functional form of the confining potential leading to the conclusion that the electromagnetic response at the edge is not universal.  In the present Reply we reexamine our field-theoretical analysis in light of these findings and (i) correct our linear response calculation and find that the numerical factor of $1/4$ in the ``shift'' term of Eq. \eqref{ourcond} should be replaced by $-1/2$, see Eq. \eqref{Re} below. This is  in agreement with Ref. \cite{Holler2016}; (ii) confirm the non-universality of the  coefficient of $p_x^2/B$ term by identifying new symmetry-allowed higher derivative edge terms in the effective action.

First, we briefly summarize the improved edge field theory derived in \cite{Moroz2015}, for details we refer to the original paper. Our system of interest is a clean quantum Hall fluid that has translation, rotation and Galilean spacetime symmetries in the bulk. Although a generic edge breaks these symmetries, a straight edge should preserve translation and Galilean symmetries along the boundary. A bosonized field theory of a chiral Luttinger liquid living at the edge of a quantum Hall fluid was introduced by Wen \cite{Wen1991a, Wen1992a} and Stone \cite{Stone1991}. For a single chiral boson $\theta$ coupled to the electromagnetic field $A_\mu$ the edge action is
\begin{equation}\label{StoneS}
S_\theta=\frac{1}{4\pi}\int d^2 x\,  \left[ \frac{1}{\nu}(D_t \theta +c D_x \theta) D_x\theta-\theta E_x\right], 
\end{equation}
where we introduced the covariant derivative
$
D_\mu\theta=\partial_\mu\theta-\nu A_\mu
$.
The edge theory can be derived from an effective description of the quantum Hall fluid in terms of statistical gauge fields with a Chern-Simons action which we will refer to as the hydrodynamic Chern-Simons model.

Importantly, we discovered in \cite{Moroz2015} that the edge theory is not invariant under Galilean boosts along a straight edge. In the bulk the electromagnetic response of a gapped abelian quantum Hall fluid is encoded in the Chern-Simons theory, which is Galilean invariant. Together this implies that the total (bulk plus boundary) action is not Galilean symmetric.  We thus concluded that for a Galilean-invariant quantum Hall liquid the theory \eqref{StoneS} is incomplete and must be improved. Building on the previous work  
\cite{Hoyos2012, Son2013},  an improved edge theory was constructed in  \cite{Moroz2015} by considering a more general problem of a quantum Hall fluid living on an arbitrary two-dimensional surface with a boundary and imposing general coordinate invariance, electromagnetic and vielbein gauge invariance. By starting from the hydrodynamic Chern-Simons model,  the resulting action for a (generically curved) edge  was found to be given by
\begin{equation}\label{StoneScov}
S_\theta=\frac{1}{4\pi}\int d^2 x\,  \left[ \frac{1}{\nu}\left(\tilde{D}_t \theta +v^x \tilde{D}_x \theta\right)\tilde{D}_x\theta-\theta \mathcal{\tilde{E}}_x \right]
\end{equation}
with
$
\tilde{D}_\mu\theta=\partial_\mu\theta-\nu \tilde{\mathcal{A}}_\mu
$.
The action \eqref{StoneScov} looks almost identical to Eq. \eqref{StoneS}, but the chiral boson in the improved theory couples to the modified gauge potential $ \mathcal{\tilde A}_\mu$ instead of the electromagnetic gauge potential $A_\mu$. In this Reply we restrict our attention only to  a quantum Hall fluid living on a flat two-dimensional surface (parametrized by Cartesian coordinates) with a flat and static edge. In this case the modified gauge potential (up to a gauge) is given by 
\beq
\begin{split}
\mathcal{\tilde A}_t&=A_t-\frac{m}{2}\delta_{ij} v^i v^j+\frac{s}{2}\epsilon^{ij} \partial_i v_j, \\
 \mathcal{ \tilde A}_i&= A_i+m v_i,
\end{split}
\eeq 
where the parameter $s$ is proportional to the ``shift'' $\mathcal{S}$ (see below) and  $v^i$ is the velocity of the quantum Hall fluid. In the present formulation, the velocity field is not independent, but is fixed by the electric and magnetic fields $E_i$ and $B$ via the Euler equation
\begin{equation} \label{Euler}
-m(\partial_t+v^k\partial_k )v_i=E_i+Bv^k\epsilon_{ki}.
\end{equation}

In \cite{Moroz2015} the effective theory of a chiral Luttinger liquid was organized according to the derivative expansion with the following power-counting scheme
\beq \label{expan}
\theta\sim \epsilon^{-2}, \quad A_i\sim \epsilon^{-1}, \quad A_t\sim \epsilon^0 \quad \partial_i \sim \epsilon, \quad \partial_t \sim \epsilon^2,
\eeq
where $\epsilon\ll 1$. This power-counting is consistent with the expansion used in \cite{Hoyos2012}.  In this derivative expansion the Lagrangian in Eq. \eqref{StoneS} is $O(\epsilon^{-1})$ and will be called the leading-order Lagrangian. On the other hand, the improved action \eqref{StoneScov} contains beyond the leading-order corrections that start with the next-to-leading $O(\epsilon)$ terms. Note also that in this power-counting the edge is smooth in the following sense: the curvature of the confining potential $\partial_y^2 A_t\sim O(\epsilon^{2})$ is parametrically small as compared to its slope $\partial_y A_t\sim O(\epsilon)$. Here $y$ is a Cartesian coordinate parametrizing the direction perpendicular to the edge.

Using the improved edge theory \eqref{StoneScov}, in \cite{Moroz2015} we computed the longitudinal conductivity \eqref{ourcond} at the edge.
We will now compare our prediction with the results found in \cite{Holler2016}. Consider first the case of a confining edge potential that is strictly linear. In the small wave vector regime the real part of the longitudinal conductivity of a non-interacting integer quantum Hall fluid was found in \cite{Holler2016} to be given by
\beq \label{Re}
\text{Re} \sigma(\omega, p_x)=\frac{\nu c}{2} \Big( 1-\frac{\mathcal{S}}{2}\frac{p_x^2}{B} \Big) \delta(\omega- c p_x),
\eeq
which has the same structure as the real part of Eq. \eqref{ourcond}, but differs from it by the sign and absolute value of the numerical coefficient that multiplies the ``shift'' $\mathcal{S}$. 
First, we find that in order to be consistent with notations used in \cite{Moroz2015}, the shift $\mathcal{S}=-2s$.\footnote{The ``shift'' $\mathcal{S}$ is defined on a sphere by $N_\phi=\nu^{-1}N-\mathcal{S}$, where $N>0$ is the particle number and $N_{\phi}$ is the magnetic flux number. Using this definition and following the arguments from Sec. 4.2 of \cite{Wen1995}, we find that the hydrodynamic Chern-Simons model Eq. (36) in \cite{Moroz2015} fixes $\mathcal{S}=-2s$. The parameter $s$ in our effective theory unfortunately differs by a sign from (i) the parameter $s$ introduced by Wen and Zee in \cite{Wen1992}; (ii) the parameter $\bar s$ commonly used in the literature \cite{Read2011, Bradlyn2012}. Our parameter $s$ is (plus) the ``mean orbital spin per particle''.} Since it is different by a sign from the relation we used in \cite{Moroz2015}, this fixes the sign difference between the real part of \eqref{ourcond} and Eq. \eqref{Re}.
Even after taking this into account, the numerical prefactors extracted from Eqs. \eqref{ourcond} and \eqref{Re} differ by a factor of $1/2$. We revised carefully our calculation and realized that in \cite{Moroz2015} we have incorrectly treated the velocity field $v^i$ as independent when deriving the consistent current as a variation of the action with respect to the gauge field (see Eq. (46) in \cite{Moroz2015}). In our construction, however, the velocity should not be considered independent, since it is a function of the electromagnetic field satisfying Eq. \eqref{Euler}. 
Taking this into account and going through the same steps as in \cite{Moroz2015}, one recovers the result that is identical   to Eq. \eqref{Re}.

The real part of the conductivity \eqref{Re} is however not a universal result for a flat quantum Hall edge because one can write additional terms in the chiral Luttinger edge action which are consistent with symmetries (translations and Galilean boosts along the edge).
As the first example, we construct a term that contributes at the next-to-leading $O(\epsilon)$ order in the derivative expansion \eqref{expan} and is invariant under (time-dependent) boundary diffeomorphisms. First, we notice that any smooth spatial boundary has an associated extrinsic curvature $K_\mu=n_i \nabla_\mu t^i$ that is constructed from the normal and tangent vectors $n^i$ and $t^i$, respectively. Up to a gauge, the extrinsic curvature $K_\mu$ coincides with the bulk spin connection evaluated at the edge \cite{Gromov2016}. As a result, $K_\mu$ does not transform as a one-form under time-dependent spatial diffeomorphisms on the boundary, but can be improved to become a one-form
$
\tilde K_t=K_t-\frac{1}{2} \varepsilon^{ij} \partial_i v_j$
and
$
\tilde K_i=K_i .
$
Due to the presence of the vorticity, the time component of the improved extrinsic curvature is non-trivial even for a flat and static boundary. 
With $\tilde K_\mu$ at hand,  we can construct an additional diffeomorphism-invariant contribution to the edge action
\beq \label{new}
\Delta S_{\theta}=\frac{\alpha}{4\pi} \int d^2x (\tilde K_t+v^x \tilde K_x) \tilde D_x \theta.
\eeq 
For a flat and static boundary this term is Galilean-invariant with respect to boosts along the edge and thus has to be included into the effective edge theory.
Also note that since Eq. \eqref{new} is invariant on its own under electromagnetic gauge transformations, the coefficient $\alpha$ cannot be determined from effective theory considerations only, but depends on the microscopic details of the edge model. Presently, the microscopic origin of this term is not understood.
Importantly, the addition \eqref{new} modifies the current and the longitudinal conductivity. The resulting real part of the conductivity has still a delta-function form \eqref{Re}, but with $\mathcal{S}\to \mathcal{S}-\alpha$. 

It was argued in \cite{Holler2016} that if  the edge mode dispersion is not strictly linear,  the real part of the conductivity is not a delta-function of frequency, but is spread in frequency with a width of order $p_x^2$ in the limit $p_x\to 0$. In particular, H\"oller and Read explicitly demonstrated that for the integer quantum Hall problem confined by a potential that has a non-zero curvature, the edge mode dispersion is not linear and the result \eqref{Re} is not valid. We note here that the correction to Eq. \eqref{Re} due to the curvature of the confining potential found in \cite{Holler2016}  is of the next-to-next-to-leading order in the expansion \eqref{expan} because as mentioned above in this expansion the curvature of the potential is parametrically small. This goes beyond the next-to-leading order effects we intended to incorporate in \cite{Moroz2015}.

In a chiral Luttinger liquid edge excitations will acquire finite width as a result of interactions between chiral bosons.
For example, one can start from the chiral Luttinger model \eqref{StoneS} and introduce into the Lagrangian interaction terms between chiral bosons such as  $\sim (D_t \theta+ c D_x \theta)^2 D_x \theta$ which is of the leading order in our derivative expansion. The interaction gives rise to a self-energy that might generate non-linear corrections to the edge dispersion and  a finite decay width of chiral edge excitations. In one-dimensional Luttinger liquids this is exactly what happens, if one goes beyond the Luttinger model and takes into account non-linear corrections to the band structure close to the Fermi points which produce interaction terms such as $(\partial_x \theta)^3$ \cite{Haldane1981, Samokhin1998, Imambekov2012}.  We emphasize that the edge model \eqref{StoneScov} derived from the general coordinate invariant hydrodynamic Chern-Simons theory has no interactions between chiral bosons and thus does not lead to a finite decay width of edge excitations. We expect, however, that a generic Galilean-invariant edge theory will contain interaction terms already at the leading order in the derivative expansion such as the one quoted above. A systematic construction of such an action will not be attempted here and is postponed to a future work.

In summary, the Comment by H\"oller and Read \cite{Holler2016} motivated us to 
reexamine the calculation of the longitudinal conductivity at the edge of a Galilean invariant quantum Hall fluid. After correcting our linear response computation \cite{Moroz2015}, the conductivity that follows from the improved theory \eqref{StoneScov} of a chiral Luttinger liquid agrees with the result found in \cite{Holler2016} in the case of a linear confining potential. As argued in \cite{Holler2016}, this result however is not universal and depends on the details of the edge theory. Here we presented additional arguments supporting this statement.

\section*{Acknowledgments:}
We acknowledge discussions with Barry Bradlyn, Andrey Gromov, Judith H\"oller and Nicholas Read. The work of Sergej Moroz is supported by the Emmy Noether Programme of  German Research Foundation (DFG) under grant No. MO 3013/1-1. This work is partially supported by the Spanish grant MINECO-16-FPA2015-63667-P. C.H is supported by the Ramon y Cajal fellowship RYC-2012-10370. LR acknowledges support by the NSF Grant DMR-1001240,  through the KITP under Grant No. NSF PHY-1125915, and by the Simons Investigator award from the Simons Foundation.

\bibliography{library}

\end{document}